\documentclass[reprint,twocolumn,superscriptaddress,showpacs,nofootinbib,notitlepage,preprintnumbers]{revtex4-2}

\usepackage{graphicx}
\usepackage{hyperref}
\usepackage[dvipsnames]{xcolor}
\usepackage{amsmath,amsthm,amssymb}
\usepackage{comment}
\usepackage{enumerate}

\renewcommand{\Re}{\mathrm{Re}\,}
\renewcommand{\Im}{\mathrm{Im}\,}

\newtheorem*{theorem*}{Theorem}

\newtheorem*{problem*}{Problem}

\usepackage[all,vshift=-1em,color=Black,angle=90,scale=3]{background}
\SetBgContents{\textsc{Draft}}
\SetBgPosition{current page.west}
\SetBgContents{}

\begin{document}
\preprint{LA-UR-23-29893}
\preprint{INT-PUB-23-044}

\title{Convex optimization and contour deformations}

\author{Scott Lawrence}
\email{scott.lawrence-1@colorado.edu}
\affiliation{Department of Physics, University of Colorado, Boulder, CO 80309, USA}
\affiliation{Los Alamos National Laboratory Theoretical Division T-2, Los Alamos, NM 87545, USA}
\author{Yukari Yamauchi}
\email{yyama122@uw.edu}
\affiliation{Institute for Nuclear Theory, University of Washington, Seattle, WA 98195, USA}

\date{\today}

\begin{abstract}
We discuss various formal aspects of contour deformations used to alleviate sign problems; most importantly, relating these contour deformations to a certain convex optimization problem. As a consequence of this connection we describe a general method for proving upper bounds on the average phase achievable by the contour deformation method. Using this method we show that Abelian lattice Yang-Mills in two spacetime dimensions possesses, for many values of the complex coupling, an exponential sign problem that cannot be removed via any contour deformation.
\end{abstract}

\maketitle

\section{Introduction}\label{sec:introduction}

Lattice Monte Carlo methods provide a unique nonperturbative view of the equilibrium properties of quantum field theories. Unfortunately, lattice calculations in various regimes are beset by sign problems, resulting in an exponential scaling in spacetime volume. The most famous cases are systems with a finite density of relativistic fermions (or, closely related, the Hubbard model away from half filling), but similar obstacles arise in many other cases, including lattice simulations of real-time dynamics.

In a lattice field theory calculation, the partition function is written as an integral over lattice configurations $\phi$ of a Boltzmann factor: $Z = \int \mathcal D \phi\, e^{-S(\phi)}$. Expectation values are obtained by differentiating the logarithm of the partition function with respect to various source terms inserted into the action, and generically have the form $\langle \mathcal O \rangle = \frac 1 Z \int \mathcal D \phi\, \mathcal O(\phi) e^{-S(\phi)}$. When the action $S(\phi)$ is real-valued, the Boltzmann factor may be interpreted as a (non-normalized) probability density. Expectation values may then be evaluated numerically by sampling with respect to the probability distribution $Z^{-1} e^{-S(\phi)}$ and computing the average of the function $\mathcal O(\phi)$ on those samples.

A sign problem arises where the action is no longer real-valued, and the Boltzmann factor is therefore not a non-negative real number. Since the Boltzmann factor cannot be interpreted as a probability, Monte Carlo methods lose any direct applicability. To salvage the situation, it is standard to make use of reweighting. Noting that
\begin{equation}
    \langle \mathcal O \rangle = 
    \frac{\int e^{-\Re S}e^{-i \Im S} \mathcal O \big/ \int e^{-\Re S}}{\int e^{-\Re S} e^{-i \Im S}\big/ \int e^{-\Re S}}
    \equiv
    \frac{\langle \mathcal O e^{-i \Im S}\rangle_Q}{\langle e^{-i \Im S}\rangle_Q}
    \text,
\end{equation}
where the dependence on fields $\phi$ has been elided for readability, we have expressed the desired expectation value as a ratio of two ``quenched'' expectation values. The quenched expectation $\langle \mathcal O \rangle_Q$ is defined as an expectation value over a probability distribution proportional to $|e^{-S}|$. This restores our ability to make use of Monte Carlo methods, at an exponential price. The denominator
\begin{equation}
    \frac{\int e^{-S}}{\int |e^{-S}|} \equiv \frac{Z}{Z_Q} \equiv \langle \sigma \rangle
\end{equation}
is a ratio of two partition functions: the physical partition function $Z$ and a ``quenched'' partition function $Z_Q$ obtained by ignoring the imaginary part of the action. As such, it is typically exponentially small in the spacetime volume of the system being studied. Resolving it from zero therefore requires exponentially many samples---this is the origin of the exponential cost of a sign problem.

Contour deformation methods (see~\cite{Alexandru:2020wrj} for a review) proceed from this point by deforming the integral over real-valued field configurations to some other contour in the complex plane. When the action is holomorphic, Cauchy's integral theorem guarantees that physical quantities do not depend on the choice of integration contour. At the same time, the average phase cannot be written as the integral of a holomorphic function, and therefore may be expected to change with the choice of contour. The central goal of different contour deformation methods is to find a contour that maximizes the average phase. In recent years a variety of machine learning approaches to this end have been explored~\cite{Alexandru:2017czx,Ohnishi:2019ljc,Kashiwa:2019lkv,Alexandru:2018fqp,Alexandru:2018ddf,Lawrence:2022afv}.

The object of this work is to explore the nature of, and provide bounds on, the ``best possible'' contour---that is, the one with maximum average phase. The core facts about contour deformations that will be established in this paper are:
\begin{itemize}
    \item On any contour which is a local minimum of the quenched partition function, the effective Boltzmann factor must have no local phase fluctuations. The phase can only change at zeros of the Boltzmann factor.
    \item The Lefschetz thimbles (historically one of the earliest contours proposed to address the QCD sign problem~\cite{Cristoforetti:2012su}) constitute the $\hbar \rightarrow 0$ limit of this optimum.
    \item For any closed differential form $\alpha$ obeying $|\alpha| \le |e^{-S} dz|$, the integral $|\int_\gamma \alpha|$ is a lower bound on the quenched partition function obtainable on \emph{any} contour in the same homology class as $\gamma$.
    \item The task of optimizing such bounds is one of convex optimization, and is dual to a convex optimization problem closely related to that of optimizing contours.
    \item There is no duality gap: the best possible bound of this form on the quenched partition function is equal to the optimal quenched partition function obtainable by contour deformation.
\end{itemize}

Finally, as an application of these results, we examine Abelian lattice Yang-Mills in two spacetime dimensions, at complex gauge coupling (also studied in~\cite{Kashiwa:2020brj}). For real-valued fields this theory, like many others, exhibits a sign problem that scales exponentially in the spacetime volume. We find that there are some values of the gauge coupling for which perfect contours exist, and the exponential sign problem can be entirely removed via contour deformation. For other ranges of the complex gauge coupling, however, we are able to show that no contour deformation can entirely remove the sign problem. For these values of the coupling we provide a rigorous, and exponentially falling, upper bound on the average sign that can be obtained via contour deformation alone.

This paper is structured as follows. In Section~\ref{sec:deformations} we outline the basic facts of contour deformations and their application to the sign problem. No results in this section are new, although the derivation in Section~\ref{sec:perfect} of the properties of locally perfect contours is crisper and more rigorous than the original argument in~\cite{Lawrence:2021izu}. Section~\ref{sec:bounds} presents the core technical innovation in this work: a general method for establishing lower bounds on the quenched partition function obtainable by contour deformation (and therefore, upper bounds on the performance of contour deformation algorithms, as measured by the average phase). We observe in Section~\ref{sec:convex-optimization} that the task of optimizing this bound is related to the task of optimizing contours via convex duality, and show that there is no duality gap. The connection between locally perfect contours and Lefschetz thimbles is covered in Section~\ref{sec:thimbles}. We apply this class of bounds to two toy models in Section~\ref{sec:toys}, and to Abelian Yang-Mills at complex coupling in~\ref{sec:yang-mills}. Finally, Section~\ref{sec:discussion} comprises a discussion of the many questions left open---or raised---by this work.

\section{Contour deformations}\label{sec:deformations}
The purpose of this section is to review the basic facts of contour deformations as applied to the sign problem, as well as to establish useful mathematical background for the rest of the paper. No result in this section is new, although the discussion of locally perfect contours is considerably improved from~\cite{Lawrence:2021izu}. The reader familiar with the recent literature on contour deformation methods may wish to skip to the section below, referring back to this section only as needed.

\subsection{Contour integrals and quenched integrals}\label{sec:integrals}
A partition function in lattice quantum field theory typically has the form
\begin{equation}\label{eq:partition}
    Z = \int d^N\!\phi\,e^{-S(\phi)}
    \text,
\end{equation}
where the exponent $S(\phi)$ is referred to as the lattice action, and the $N$ variables $\phi$ being integrated over represent a field configuration.

The integration in Eq.~(\ref{eq:partition}) is implicitly taken\footnote{In the case of periodic variables $\phi$---including both concrete examples discussed in Section~\ref{sec:toys} below---the original domain of integration is the $N$-torus $(S^1)^N$, and the complexified domain is topologically $(S^1)^N \times \mathbb R^N$.} to be over $\mathbb R^N$. In the case of a sign problem, the action is a function from the space $\mathbb R^N$ of field configurations to the complex numbers. The first step to applying contour deformation methods to treat this sign problem is to analytically continue the action, so that $e^{-S}$ is a holomorphic\footnote{For bosonic theories, the action $S$ itself is typically holomorphic; when fermions have been integrated out, zeros of the fermion determinant yield logarithmic singularities in the action.} function from $\mathbb C^N$ to the complexes. For the purpose of this work, we will assume that this analytic continuation is automatic, and we will not distinguish between the original action and the analytically continued one.

With the analytic continuation complete, the partition function is now re-written as a contour integral over $\mathbb R^N \subset \mathbb C^N$. Cauchy's integral theorem guarantees that continuous deformations of the contour do not change the integral, although they can affect the sign problem. As a result, physical measurements do not change, but the performance of numerical algorithms may be improved. The central goal of contour deformation-based methods is to find contours with favorable algorithmic properties.

This procedure is best made precise in the language of differential forms\footnote{An excellent introduction and reference for differential forms, for physicists, is found in~\cite{arnol2013mathematical}. Only the minimal facts will be reviewed here, with an emphasis on the specialization to contour integration.}. The underlying theorem that enables the contour deformation is Stokes' theorem: for any region $\Omega$ and differential form $\omega$, we have
\begin{equation}
    \int_\Omega d\omega = \int_{\partial \Omega} \omega\text.
\end{equation}

The Boltzmann factor is used to construct a differential form $\omega \equiv e^{-S} \bigwedge_i d z_i$. For a holomorphic Boltzmann factor, it is straightforward to see that $\omega$ is closed:
\begin{equation}
    d \omega = \sum_j \frac{\partial e^{-S}}{\partial z} d z_j \wedge \left[\bigwedge_i dz_i\right] = 0
\end{equation}
As a result, the integral along the boundary of any region must vanish. This defines the `permitted' contour deformations: they differ from the original contour only by the boundary of some region $\Omega$, where $d\omega = 0$ holds at least in $\Omega$.

To discuss the quenched partition function (and therefore the average phase), we need a notion of the absolute value of a differential form. This is easy enough: recall that a differential form is a function from the tangent space to the complex numbers. We define
\begin{equation}
    |\omega_\phi|(v_1,\ldots,v_k) = |\omega_\phi(v_1,\ldots,v_k)|
    \text.
\end{equation}
The object $|\omega|$ may be integrated in the usual way, defining the quenched partition function $Z_Q^{(\gamma)} = \int_\gamma |\omega|$. However, $|\omega|$ is not a differential form, as it is neither linear nor antisymmetric in its arguments. We will refer to $|\omega|$ as an \emph{absolute differential form}\footnote{This terminology is from~\cite{nlab:absolute_differential_form} and is particularly suggestive in our context. Another name for the same object is an \emph{even $k$-density}~\cite{gel1977nonlocal}.}. Absolute differential forms have the following two properties of note:
\begin{itemize}
	\item If any argument is scaled $v_i \mapsto \lambda v_i$, then the value of the absolute differential form is scaled by $|\lambda|$, and
	\item The value of the absolute differential form is unaffected by the addition to one vector of a linear combination of the other vectors\footnote{The same is true for ordinary differential forms, from which absolute differential forms inherit this property.}. That is, $|\omega|(v_1, \ldots) = |\omega|(v_1 + \sum_{i>1} c_i v_i, \ldots)$.
\end{itemize}
These properties in turn imply that absolute differential forms are invariant under permutation of the arguments. As a result, the integral $\int_\gamma |\omega|$ over a contour $\gamma$ is well-defined, and does not depend on the orientation of $\gamma$.

In a numerical computation, the contour $\gamma$ is always parameterized by the real plane $\mathbb R^N$. That is, a function $\tilde \phi(\phi) : \mathbb R^N \rightarrow \mathbb C^N$ is chosen with image $\gamma$. The effective action is then defined according to
\begin{equation}
    S_{\mathrm{eff}}^{[\tilde \phi]}(\phi) = S\big(\tilde\phi(\phi)\big) - \log \det \frac{\partial\tilde\phi(\phi)}{\partial \phi}
    \text.
\end{equation}
The partition function and quenched partition function may now be respectively expressed as
\begin{equation}
    Z = \int d^N\!\phi\,e^{-S_{\mathrm{eff}}(\phi)}
    \text{ and }
    Z_Q = \int d^N\!\phi\,\left|e^{-S_{\mathrm{eff}}(\phi)}\right|
    \text.
\end{equation}
This approach has the practical advantage of being concrete and amenable to numerical (especially Monte Carlo) methods. However, it is often unwieldy, and the results in this paper are more easily expressed in the language of differential (and absolute) forms. It also has the more serious shortcoming of excluding \emph{a priori} any contours which are not homeomorphic to $\mathbb R^N$.

\subsection{Perfect and locally perfect contours}\label{sec:perfect}

We want to maximize the average phase $\langle \sigma\rangle \equiv \frac{Z}{Z_Q}$. As we are performing this maximization over a class of integration contours on which the partition function $Z$ is constant, this is equivalent to minimizing the quenched partition function over the space of contours.

The most desirable property for a contour $\gamma$ to have is that the average phase be exactly $1$; that is, the quenched partition function $Z_Q^{(\gamma)}$ should be equal to the absolute value $|Z|$ of the partition function. Any such contour represents a global minimum of the quenched partition function. The converse is not true; the global minimum of the quenched partition function need not achieve $Z_Q = |Z|$. In such a case we would say that there is no perfect contour.

In general, studying the global minima of a function is difficult. In this case there is little we can say, although as shown in Appendix~\ref{app:nonunique}, it can be the case that there are several distinct perfect contours. For now, let us settle for examining the properties of local minima of $Z_Q$. Parameterizing the contour---at least locally---by $z(\xi)$, we will take the functional derivative of $Z_Q$ with respect to $z(\xi)$ and examine the requirement that this vanish.

A bit of care is required in defining this derivative. First consider the case of a real-valued function $F(z)$ over the complex plane. The correct condition on the derivatives of $F$ for identifying minima (in general, saddle points) is that
\begin{equation}
    0 = \left(\frac{d}{dx} + i \frac{d}{d y}\right) F(x+iy) \equiv \frac{D}{D z} F(z)
    \text.
\end{equation}
Note that the gradient operator used here is not the holomorphic (or Wirtinger) derivative. This equation guarantees that the derivatives of $F$ with respect to the real and imaginary parts of $z$ separately vanish.

When $F$ happens to be the real part of a holomorphic function $F(z) = \Re f(z)$, this gradient operator may be evaluated in terms of the Wirtinger derivative via
\begin{equation}
    \frac{D}{D z} \Re f(z) = \overline{\frac{\partial f}{\partial z}}
    \text,
\end{equation}
providing a connection to the holomorphic derivative of $f$ via conjugation. In particular, as long as $f$ is holomorphic, the condition $0 = \frac{\partial f}{\partial z}$ identifies saddle points.

Similarly, the appropriate saddle point condition on $Z_Q$ is
\begin{equation}\label{eq:ZQ-saddle}
    0 = \frac{\Delta}{\Delta z(\xi)} Z_Q\text.
\end{equation}
By analogy with the case of the ordinary gradient above, the functional gradient $\frac{\Delta}{\Delta z(\xi)}$ has the following defining property: for a real-valued functional $G[z(\xi)]$ that happens to be the real part of a holomorphic functional $g[z(\xi)]$, we have
\begin{equation}
    \frac{\Delta}{\Delta z(\xi)} \Re g[z(\xi)]
    =
    \overline{\frac{\delta g}{\delta z(\xi))}}
    \text.
\end{equation}

Beginning from the saddle-point condition Eq.~(\ref{eq:ZQ-saddle}), and applying the above identity, we find
\begin{align}
    0 = \int d \xi'
    &e^{-\Re S[z(\xi')]} \left|J(\xi')\right| \times \nonumber\\
    &\left[
     S[z(\xi')-\frac{\delta}{\delta z(\xi)} \left(\log \det J(\xi')]\right)
    \right]^*
    \text.
\end{align}
Here $J$ denotes the Jacobian matrix $J_{ij}(\xi) = \frac{\partial z_i(\xi)}{\partial \xi_j}$.
Differentiating the logarithm, noting that $\overline{\delta(\xi-\xi')} = \delta(\xi-\xi')$, and integrating by parts immediately yields
\begin{equation}
    0 =
    e^{-\Re S_{\mathrm{eff}}}
    \left[
    \overline{\frac{\partial S}{\partial z_i}}
    +
    \frac{\partial\overline{J}_{ji}^{-1}}{\partial \xi_j}
    -\overline{J}_{ji}^{-1}
    \frac{\partial}{\partial \xi_j} \Re S_{\mathrm{eff}}
    \right]
    \text,
\end{equation}
which must be satisfied for all $\xi$ and indices $i$. The index $j$ is implicitly summed over.

This expression is difficult to interpret meaningfully. A primary reason for this is that it contains derivatives with respect to both $z$ (the complex coordinate of the contour) and $\xi$ (the parameterizing coordinate). We may address this by multiplying the whole expression by the Jacobian $J_{ik}$, with the now-internal index $i$ summed over. Simplifying appearances of $J^{-1} J = I$, we obtain
\begin{equation}
    0 =
    e^{-\Re S_{\mathrm{eff}}}
    \left[
    \frac{\partial S}{\partial z_i} J_{ik}
    -
    J_{ji}^{-1} \frac{\partial J_{ik}}{\partial \xi_j}
    -
    \frac{\partial}{\partial \xi_k} \Re S_{\mathrm{eff}}
    \right]
    \text.
\end{equation}
The combination of the first two terms is readily verified to be none other than the derivative of the effective action with respect to $\xi_k$. As a result, we finally obtain the following equation which must be satisfied by any local minimum of $Z_Q$ (and is in fact sufficient to guarantee that $Z_Q$ is locally minimized):
\begin{equation}
    0 =
    e^{-\Re S_{\mathrm{eff}}}
    \frac{\partial}{\partial \xi_k} \Im S_{\mathrm{eff}}\text.
\end{equation}

We conclude that any local minimum (and in fact, any local maximum or saddle point) of the quenched partition function is a contour with no local phase fluctuations\footnote{A much different argument for the same claim appeared first in~\cite{Lawrence:2021izu}; the argument here is more rigorous.}. The phase of the integrand $\omega$ on such a contour can change only at points where that phase is undefined---in other words, where the Boltzmann factor vanishes. The contour therefore consists of the disjoint union of individual pieces, each of which has no sign problem, but which potentially have differing phases, and are separated by zeros of the Boltzmann factor (which may be at infinity). We term this situation a \emph{global sign problem}, distinguished from the \emph{local sign problem} when $\partial \Im S \ne 0$. In short: local minima of $Z_Q$ may have a global sign problem, but no local sign problem---we term such contours \emph{locally perfect}.

Note that these pieces are not Lefschetz thimbles. On Lefschetz thimbles, the Boltzmann factor has constant phase but the complex measure $dz$ generically has local fluctuations. On a locally perfect contour, the Boltzmann factor $e^{-S}$ has local phase fluctuations which are exactly cancelled by those of the complex measure. Further discussion of the relation to Lefschetz thimbles is in Section~\ref{sec:thimbles}.

\subsection{Analytically continued normalizing flows}\label{sec:nf}

In a few cases, perfect contours can be obtained by analytic continuation of normalizing flows~\cite{Lawrence:2021izu}. This perspective provides evidence for the existence of globally perfect contours in certain cases where the sign problem is mild. 

A normalizing flow is an invertible map $z(x)$ from $\mathbb R^N$ to some target space (here assumed, for convenience, to be $\mathbb R^N$ as well), such that the normal distribution on the domain induces a desired probability distribution $p(z)$ on the target space:
\begin{equation}\label{eq:nf}
    p(z(x)) \det \frac{\partial z}{\partial x} = e^{-x^2 / 2}
    \text.
\end{equation}
A standard theorem asserts that for any probability distribution on $\mathbb R^N$, an exact normalizing flow exists~\cite{villani2021topics}. In practice, for many numerical applications it is useful to find maps $z(x)$ which only approximately satisfy Eq.~(\ref{eq:nf}). This is an increasingly common technique in the context of lattice field theory for the purpose of accelerating sign-free Monte Carlo calculations; e.g.~\cite{Albergo:2019eim,Abbott:2023thq}.

Of course, any map $z(x)$ is a normalizing flow for some probability distribution, readily calculated from Eq.~(\ref{eq:nf}). We might then generalize the notion of a normalizing flow by allowing the map $z(x)$ to be complex valued. The resulting map---termed a \emph{complex normalizing flow}---serves to parameterize an integration contour $\gamma \subset \mathbb C^N$ by the real plane, and in fact defines a probability distribution on this contour. As a result, we see that a complex normalizing flow provides a (globally) perfect integration contour for some distribution $p$ which is complex-valued on the real plane.

Physical Boltzmann factors $p(z)$ of interest also implicitly depend on some generalized coupling constants which we will label $\lambda$. Suppose that for real $\lambda$, $p(z; \lambda)$ is a (non-negative real-valued) probability distribution on field configurations $z$, but when $\Im \lambda \ne 0$, we encounter a sign problem. For values of $\lambda$ sufficiently close to the real line, we can find a perfect contour for such a sign problem as follows. We find a family of normalizing flows $z(x;\lambda)$, such that $z$ is an analytic function\footnote{Note that our ability to do this is a stronger statement than the standard theorem of existence, and is here only a conjecture.} of $\lambda$. This family of normalizing flows may be analytically continued some finite amount off of the line of real $\lambda$, and where this is possible, $z(x;\lambda)$ induces a perfect integration contour for $p(z;\lambda)$.

This perspective will be further explored in Section~\ref{sec:cosine} below, in the case of a one-dimensional integral where exact information about both the normalizing flows and the available contour deformations is available.

\section{Quenched bounds}\label{sec:bounds}

For some lattice actions, it can be proven that no perfect contour exists. An extreme case was discussed in~\cite{Lawrence:2022afv}, motivated by the existence of partition function zeros on the plane of complex lattice couplings. Any non-zero lower bound on the quenched partition function near these points implies that, sufficiently close to the zero of the partition function, there is no perfect contour (and in fact the best-possible average phase approaches $0$).

In general, because the partition function does not depend on the choice of contour, any lower bound on the quenched partition function translates directly into an upper bound on the average phase $\langle \sigma\rangle = \frac{Z}{Z_Q}$. The purpose of this section is to detail one method by which such bounds may be obtained.

One lower bound on the quenched partition function is well-known: $Z_Q \ge |Z|$. This follows from the observation that, for any differential form $\omega$,
\begin{equation}
\left|\int_\gamma \omega\right| \le \int_\gamma | \omega|\text.
\end{equation}
Note that the usefulness of this inequality stems from the fact that the right-hand side depends on the choice of integration contour $\gamma$, while the left-hand side is the same for any homologous contour.

We can potentially improve this bound by finding another closed differential form $\alpha$ satisfying $|\alpha| < |\omega|$. Now we have a short series of inequalities:
\begin{equation}
    \left|\int_\gamma \alpha\right|
    \le \int_\gamma |\alpha|
    \le \int_\gamma |\omega|
    = Z_Q
    \text.
\end{equation}
Once again, because $\alpha$ is closed, the first expression does not depend on the choice of contour $\gamma$, as long as $\gamma$ lies in the physical homology class. For any closed form $\alpha$ obeying $|\alpha| < |\omega|$, the absolute value of $\int \alpha$ serves as a lower bound on the quenched partition function.

It is up to the user of this method to find a form $\alpha$ that gives a useful lower bound on the quenched partition function (and therefore a useful upper bound on the achievable average phase). One piece of guidance may be useful. In physically motivated examples, the form $\omega$ is proportional to $dz$, or a wedge product $\bigwedge_i dz_i$:
\begin{equation}
    \omega = e^{-S(z)} dz
    \text.
\end{equation}
In these cases, no holomorphic $\alpha$ will yield a nontrivial bound: $\alpha$ must have some direct dependence on $\bar z$. To see this, note that a holomorphic $\alpha$ may be written $\alpha = g dz$, and the inequality $|\alpha| \le |\omega|$ implies that $|g| \le |f|$. But both $f$ and $g$ are holomorphic, and so the quotient---a meromorphic function---obeys $| g / f | \le 1$. The only meromorphic functions with bounded magnitude are constant, so $\alpha$ is proportional to $\omega$ and does not yield a stronger bound.

\section{Convex optimization}\label{sec:convex-optimization}

A convex optimization problem\footnote{A canonical introduction and reference for convex optimization is~\cite{boyd2004convex}; we will repeat from there the minimum background necessary for the results of this work.} is one that can be written in the following form:
\begin{align}
    \text{minimize } f(x) \text{ subject to }&x \in X\nonumber\\
    \text{and }& c_i = g_i(x)\nonumber
\end{align}
where $X$ is a convex set, $f$ a convex, real-valued function on $X$, and the functions $g_i$ are linear. Note that any convex constraints, often written $0 \le h(x)$ with $h$ a convex function, have been implicitly included in the definition of $X$.

In this section we will show that the problem of optimizing a bound of the form discussed in the previous section is an instance of convex optimization. Similarly, there exists a mild generalization of the contour optimization task which is also a convex optimization problem.

These two optimization problems are in fact dual to each other, in a precise sense. When two convex optimization problems are dual, any feasible point of the dual problem---that is, any point which obeys all constraints, but may not be optimal---is necessarily a lower bound on the set of feasible points of the original (or \emph{primal)} problem, and therefore is a lower bound on the solution to the primal problem. When the maximum of the dual problem is equal to the minimum of the primal problem, the problem is said to exhibit \emph{strong duality}.

This duality and its consequences are the subject of Subsections~\ref{sec:duality} and~\ref{sec:strong}. In particular, we will show that one can always find a tight bound of the form considered in the previous section.

\subsection{Convexity}\label{sec:convexity}

Denote by $\mathfrak A_\omega$ the space of closed differential forms $\alpha$ obeying $|\alpha| \le |\omega|$. We will first show that this space is convex, as follows. For any two forms $\alpha_1,\alpha_2 \in \mathfrak A_\omega$, the linear combination $c_1\alpha_1 + c_2\alpha_2$ is closed:
\begin{equation}
    d(c_1\alpha_1 + c_2\alpha_2) = c_1 d\alpha_1 + c_2 d\alpha_2 = 0
    \text,
\end{equation}
Moreover, as long as $c_1 + c_2 = 1$, this linear combination is also dominated by $|\omega|$:
\begin{equation}
    |c_1\alpha_1 + c_2\alpha_2| \le c_1 |\alpha_1| + c_2 |\alpha_2| \le (c_1 + c_2)|\omega|
    \text.
\end{equation}
In the previous section we showed that given a differential form $\alpha \in \mathfrak A_\omega$, we obtain a bound on the quenched partition function $Z_Q(\gamma) = \int_\gamma |\omega| \ge \int_{\gamma_0} |\alpha|$, where $\gamma_0$ is the original contour of integration\footnote{In typical circumstances, this contour is simply $\mathbb R^N \subset \mathbb C^N$.
} and $\gamma$ is any integration contour in the same homology class. The task of finding the tightest such bound is equivalent to the task of maximizing $|\int_{\gamma_0} \alpha|$ over $\alpha \in \mathfrak A_\omega$. This objective function is of course convex. Alternatively, we might restrict to forms $\alpha$ obeying $\int \alpha \in \mathbb R$, and then maximize (or minimize) $\int \alpha$ itself. Either way, we see that the task of optimizing the bound is one of convex optimization. To be concrete, the optimization problem is:
\begin{align}\label{eq:opt-bounds}
    \text{maximize } \left|\int_{\gamma_0} \alpha \right| \text{ subject to }&d\alpha = 0\nonumber\\
    \text{and }&|\alpha| \le |\omega|\text.
\end{align}

Now we move from optimizing bounds on contours to optimizing the contours themselves. We must begin by constructing a generalization of the notion of a contour of integration, which we will term (in a slight abuse of terminology\footnote{See~\cite{HatcherAT} for the definition of chain groups in singular homology, which is very similar.}) an $N$-chain. An $N$-chain is a linear combination of (perhaps infinitely many) $N$-contours. Such linear combinations may in principle be taken with integer, real, or complex coefficients; in this paper we will focus on the case of real coefficients. Integration on an $N$-chain $\sum c_i \Gamma_i$ is defined via linearity:
\begin{equation}
	\int_\Gamma \alpha = \sum_i c_i \int_{\Gamma_i}\alpha\text.
\end{equation}

Equivalent definitions are possible, and one in particular will prove useful later. Recall that the \emph{Grassmannian} associated to a point $x$ is the space of $N$-dimensional linear subspaces of the tangent space at $x$. An $N$-chain associates to each point $x$ and each element of its Grassmannian, a real number.

We introduced in Section~\ref{sec:integrals} the notion of an absolute differential form. Part of the importance of $N$-chains is that, via integration, they may be viewed as linear functions of the space of absolute differential forms, and in fact every such linear function may be written as an $N$-chain.

Consider the space $C_N(\mathbb C^N; \mathbb R)$ of $N$-chains on complex field space, with coefficients in $\mathbb R$. Any integration contour is of course an element of $C_N(\mathbb C^N; \mathbb R)$, but not all such $N$-chains correspond to sensible integration contours. The integration contours, as usually understood, are constructed only with integer coefficients, and so that space is written $C_N(\mathbb C^N;\mathbb Z)$.

Now, this space of $N$-chains with real coefficients is clearly a linear space. However, most $N$-chains aren't relevant to us, as they lie in a different homology class from $\mathbb R^N$. Define $\mathcal C(\mathbb R)$ to be the subset of $C_N(\mathbb C^N; \mathbb R)$ that is in the same homology class as the ``physical'' contour $\gamma_0$ used to define the partition function $Z = \int_{\gamma_0} \omega$. This is also an affine constraint defining a convex subspace: given two chains $\Gamma_1$ and $\Gamma_2$ in this homology class, any linear combination $c_1 \Gamma_1 + c_2 \Gamma_2$ obeying $c_1 + c_2 = 1$ is also in the physical homology class.

Integration is defined on $\mathcal C(\mathbb R)$, not just on its subset with integer coefficients. Thus, we can define the quenched Boltzmann factor for each chain $\Gamma \in \mathcal C(\mathbb R)$ in the usual way, via $Z_Q^{(\Gamma)} = \int_\Gamma |\omega|$. This is manifestly a convex function on $C_N(\mathbb C^N; \mathbb R)$, and therefore on the affine subspace $\mathcal C(\mathbb R)$.

We have shown that finding an optimal element of $\mathcal C(\mathbb R)$ minimizing the quenched partition function is a convex optimization task:
\begin{equation}\label{eq:opt-chains}
	\text{minimize }\int_\Gamma |\omega| \text{ subject to } [\Gamma] = [\gamma_0]
\end{equation}
This is \emph{not} the same task as optimizing a contour, as the valid integration contours form a non-convex subset of $\mathcal C(\mathbb R)$. We will refer to members of $\mathcal C(\mathbb R)$ as generalized contours\footnote{The term ``fuzzy contour'' has also been suggested~\cite{bedaque:private}.}, and (\ref{eq:opt-chains}) as the generalized contour optimization problem.

\subsection{Duality}\label{sec:duality}

Above we presented two convex optimization problems: one for optimizing the bound via choice of differential form $\alpha$, and one for optimizing the generalized contour. In fact these problems are related, as we will now see.

To begin let us review convex optimization in the classical case, following the exposition in~\cite{boyd2004convex}. Consider an optimization task of the form
\begin{equation}\label{eq:primal-problem}
    \text{minimize }f_0(x)\text{ subject to }f_i(x) \le 0\text,
\end{equation}
where the inequality must be satisfied for each of several convex functions $f_i$ (not to include $f_0$), and any linear constraints on $x$ have been implicitly included, for brevity of the below discussion, into the space over which $x$ is to be optimized. The function $f_0$ is also presumed to be convex\footnote{We assume that the functions $f$ are convex for brevity. Many of the results below do not require this, while a few do; the interested reader should consult a source dedicated to convex optimization rather than attempt to generalize from this discussion.}.

We may construct from such an optimization problem a \emph{Lagrangian function}
\begin{equation}
    L(x,\lambda) = f_0(x) + \sum_i \lambda_i f_i(x)
    \text,
\end{equation}
a \emph{Lagrange dual function}
\begin{equation}
    g(\lambda) = \inf_x L(x,\lambda)
    \text,
\end{equation}
and finally a convex optimization problem termed the \emph{Lagrange dual problem}:
\begin{equation}\label{eq:lagrange-dual-problem}
    \text{maximize }g(\lambda)\text{ subject to }\lambda_i \ge 0
    \text.
\end{equation}

Note that the solution to the Lagrange dual problem (\ref{eq:lagrange-dual-problem}) is given by $d^* = \max_\lambda \inf_x L(x,\lambda)$. Less trivially, the solution to the primal problem (\ref{eq:primal-problem}) is given by $p^* = \inf_x \max_\lambda L(x,\lambda)$. The first and most important result is the statement of \emph{weak duality}:
\begin{equation}\label{eq:weak-duality}
    d^* \le p^*\text.
\end{equation}
The optimum of the dual problem is a lower bound on that of the primal (original) problem.

In the case of convex optimization, the inequality of Eq.~(\ref{eq:weak-duality}) is often tight. A sufficient condition for this to be true is that there is a point $x$ which is strictly feasible---that is, for which $f_i(x) < 0$. This is termed the Slater condition.

One additional fact about convex duality provides useful intuition: the concept of \emph{complementary slackness}. Each Lagrange multiplier $\lambda_i$ corresponds to a convex constraint $f_i(x)$. It may be shown that, when computing $d^* = \sup_\lambda \inf_x L(x,\lambda)$, it will always be the case that at the optimal pair $(x,\lambda)$, we have $\lambda_i f_i(x) = 0$ for each $i$. Thus, for each constraint, $\lambda_i = 0$ must hold unless the constraint is ``active'' at the optimum.

We will now apply this theory to the optimization of bounds (\ref{eq:opt-bounds}) above, and construct the Lagrange dual problem.

The problem of optimizing $\alpha$ is a convex maximization problem. Let us now construct the Lagrange dual problem, which will be a convex minimization problem. First we must pass from the classical case to the infinite-dimensional case. It is convenient to write the optimization problem (\ref{eq:opt-bounds}) in a slightly different form, to mimic (\ref{eq:primal-problem}) as closely as possible:
\begin{equation}
    \text{minimize }-\left|\int_{\gamma_0} \alpha\right|\text{ subject to }|\alpha| - |\omega| \le 0\text.
\end{equation}
The sum over Lagrange multiplier terms is now replaced by an integral over a chain $\Gamma$ of the constraint functions. It is important to verify that the chain $\Gamma$ contains exactly one real number for each independent constraint, as follows. At a fixed point, if the constraint function $|\alpha|-|\omega|$ is known on a single sequence of $N$ vectors $v_1,\ldots,v_N$, then by the properties of absolute differential forms discussed in Section~\ref{sec:integrals}, the value of $|\alpha|-|\omega|$ is known on all sequences of vectors in the linear subspace spanned by $v_1,\ldots,v_N$. Thus the only independent constraints are those associated to different linear subspaces.

The resulting Lagrangian function reads
\begin{equation}
    \mathcal L(\alpha,\Gamma) = - \left|\int_{\gamma_0} \alpha \right| + \int_\Gamma |\alpha| - \int_\Gamma |\omega|\text.
\end{equation}
The Lagrange dual function is then the infimum of this Lagrangian over all closed forms $\alpha$:
\begin{equation}
\mathcal{G}(\Gamma) = \inf_{\alpha} \mathcal L(\alpha,\Gamma)
= \inf_\alpha \left[\int_\Gamma |\alpha| - \left|\int_{\gamma_0} \alpha\right| \right] - \int_\Gamma |\omega|
    \text.
\end{equation}
Finally we define the Lagrange dual problem. In the classical case, we require a constraint $\lambda_i \ge 0$. Here this is not necessary, as the sign (in general, orientation) of $\Gamma$ does not affect quenched integrals. Alternatively, one might ``impose'' the constraint that for any differential form $\nu$, we have $\int_\Gamma |\nu| \ge 0$. Either way, the dual problem simply reads:
\begin{equation}\label{eq:dual-initial}
    \text{maximize }\mathcal{G}(\Gamma)
    \text.
\end{equation}

To interpret this dual problem, we note that many chains $\Gamma$ are ``forbidden'' by the fact that $\mathcal{G}(\Gamma) = -\inf$. We are therefore restricted to chains $\Gamma$ such that, for any closed form $\alpha$, we have
\begin{equation}\label{eq:feasible-Gamma}
    \int_\Gamma |\alpha| \ge \left|\int_{\gamma_0}\alpha\right| \text.
\end{equation}
Conceptually, this is requiring that $\Gamma$ be ``at least as large'' as the fiducial integration contour $\gamma_0$. This constraint is automatically obeyed by any chain in the same homology class as $\gamma_0$, as for any such chain we have the usual chain of inequalities:
\begin{equation}
	\int_\Gamma |\alpha| \ge \left|\int_\Gamma \alpha \right| = \left|\int_{\gamma_0} \alpha \right|\text.
\end{equation}
However these are not the only chains that satisfy (\ref{eq:feasible-Gamma}). As an example, for any chain $\Gamma$ in the same homology class as $\gamma_0$, and any real $c$ obeying $|c| \ge 1$, $c \Gamma$ also obeys (\ref{eq:feasible-Gamma}). Beginning again with a chain $\Gamma$ in the correct homology class, we can also add arbitrary disjoint chains without decreasing the integral of $|\alpha|$. Therefore, as written, the optimization problem (\ref{eq:dual-initial}) is over a larger space than desired.

We can modify (\ref{eq:dual-initial}) to exclude such chains without changing the optimum. For any $\Gamma$ written as a disjoint sum of some chain $\Gamma_0$ in the physical homology class, and an additional chain $\tilde\Gamma$, we can remove the contribution of $\tilde\Gamma$ while decreasing the quenched integral of $|\omega|$. Therefore no such chain can be the optimum of (\ref{eq:dual-initial})---it is always preferable to simply have $\Gamma_0$.

Finally we are able to rewrite the optimization problem (\ref{eq:dual-initial}) in a more desirable form, as an optimization over chains in the physical homology class:
\begin{equation}
	\text{minimize} \int_\Gamma |\alpha| \text{ subject to } [\Gamma] = [\gamma_0]\text.
\end{equation}

\subsection{Strong duality}\label{sec:strong}

It may easily be seen that Slater's condition does not hold for (\ref{eq:opt-bounds}): for typical $\omega$ there is some point such that $|\omega| = 0$, and as a result there can be no strictly feasible form $\alpha$. Fortunately, Slater's condition is merely a sufficient condition for strong duality, and not a necessary one.

As it happens, a small change to the above optimization problems restores Slater's condition. At any point $x_0$ where $\omega = 0$, we remove the constraint $|\alpha(x_0)| \le |\omega(x_0)|$ and replace it with an explicit equality $\alpha(x_0) = 0$. With this modification $\alpha(\cdot) = 0$ is strictly feasible, and Slater's condition holds.

In general such a modification might change the Lagrangian and therefore the dual problem. In this case, however, because the set of points at which $\omega = 0$ has measure $0$, the Lagrangian cannot `detect' our modification, and strong duality between the two problems above is established.

The strong duality of the two optimization problems described above may also be proven via Sion's minimax theorem~\cite{sion1958general}. A slightly weakened form of this theorem (taken from~\cite{komiya1988elementary}) states
\begin{theorem*}[Sion's minimax theorem]
Let $X$ be a compact convex subset of a linear topological space, and $Y$ a convex subset of a linear topological space. Let $f$ be a real-valued continuous function on $X \times Y$ such that
\begin{enumerate}[(i)]
    \item $f(x,\cdot)$ is concave on $Y$ for all $x \in X$, and
    \item $f(\cdot,y)$ is convex on $X$ for each $y \in Y$.
\end{enumerate}
It follows that
\begin{equation}
    \min_{x \in X} \sup_{y \in Y} f(x,y) = \sup_{y \in Y} \min_{x \in X} f(x,y)\text.
\end{equation}
\end{theorem*}

The application to the strong duality of generalized contour optimization and form optimization proceeds as follows. Recall that $\mathfrak A_\omega$ is the space of closed differential forms $\alpha$ obeying $|\alpha| \le |\omega|$---this is a compact convex subset of a linear topological space. The space $\mathcal C(\mathbb R)$ of $N$-chains is not compact, but is a convex subset of a (different) linear topological space. The Lagrangian function $\mathcal L(\alpha, \Gamma)$ is manifestly a real-valued continuous function on $\mathfrak A_\omega \times \mathcal C(\mathbb R)$. The concave/convex conditions are easily verified, and we conclude that
\begin{equation}
    \max_{\alpha\in\mathfrak A_\omega} \inf_{\Gamma\in\mathcal C_N(\mathbb C^N;\mathbb R)} \mathcal L(\alpha,\Gamma)= \inf_{\Gamma\in\mathcal C_N(\mathbb C^N;\mathbb R)} \max_{\alpha\in\mathfrak A_\omega} \mathcal L(\alpha,\Gamma)
    \text.
\end{equation}
The left-hand side represents the task of optimizing the differential form, and the right-hand side that of optimizing the generalized contour. We conclude that there is no duality gap.

We have shown that the bound on the best possible generalized contour (that is, element of $\mathcal C(\mathbb R)$) is tight. This implies that the method of the previous section provides a tight bound on the best possible contour (element of $\mathcal C(\mathbb Z)$) as long as, for every generalized contour $\Gamma$, there exists a contour $\gamma$ such that $Z_Q(\gamma) \le Z_Q(\Gamma)$. This follows from the fact that $Z_Q(\Gamma)$ is linear in $\Gamma$.

\section{Lefschetz Thimbles}\label{sec:thimbles}

The purpose of this section is to explore the relation between the perfect (and locally perfect) contours that are the central object of this paper, and the Lefschetz thimbles. In particular, we will show that in the $\hbar \rightarrow 0$ limit the two coincide.

Let us consider how a locally perfect contour changes as a function of $\hbar$. Although factors of $\hbar$ have been implicit up until this point, the quenched partition function reads
\begin{equation}
	Z_Q^{(\gamma)}(\hbar) = \int_\gamma \left|e^{-S(x) / \hbar}\right| d^Nx
\end{equation}
and we will take $\gamma_T(\hbar)$ to be the global minimum of $Z_Q$. Let us further assume that $\gamma_T(\hbar)$ is in fact a globally perfect contour at all $\hbar$ being considered, so that $Z_Q^{(\gamma)}(\hbar) = |Z(\hbar)|$. This is done without loss of generality: the case of a contour with a global sign problem can be understood by considering how each individual piece behaves.

First, what happens to $\gamma_T$ as $\hbar \rightarrow \infty$? When we parameterize $\gamma_T$ by the real plane, via a function $z_T(x; \hbar)$, the effective action reads
\begin{equation}
    S_{\mathrm{eff},T}(x;\hbar)
    = S(z_T(x;\hbar)) - \hbar \log \det J
    \text.
\end{equation}
In the limit of large $\hbar$, in order for the Boltzmann factor to be of constant phase, the imaginary part of $\log \det J$ must be exactly constant. In the case of one complex dimension, this implies that the perfect contour is in fact an affine subspace of $\mathbb C$. In more dimensions, this need not be true: there are non-flat contours on which $\Im \log \det J$ is nonetheless constant.

Next we consider the limiting case of $\hbar \rightarrow 0$ (with the limit taken along $\hbar \in \mathbb R_+$). The trade-off between the Boltzmann factor and the phase from the integration measure ($dz_1 \wedge \cdots \wedge dz_N$) is reversed from what it was before: the limiting contour must obey $\Im S = \text{const}$. In fact this property is satisfied, albeit not uniquely, by a well known set of integration contours: the Lefschetz thimbles\footnote{See \cite{Witten:2010cx} for a more detailed exposition.}. Each Lefschetz thimble extends from a saddle point of the action---that is, a field configuration $\phi_0$ obeying $\frac{\partial S}{\partial \phi}\Big|_{\phi=\phi_0} = 0$---and is defined as the union of solutions $\phi(t)$ to the differential equation
\begin{equation}
    \frac{d \phi}{d t} = \overline{\frac{\partial S}{\partial \phi}}
\end{equation}
such that $\phi(t)$ asymptotically approaches the saddle point at early times:
\begin{equation}
\lim_{t \rightarrow \infty^-} \phi(t) = \phi_0
    \text.
\end{equation}

Therefore, any Lefschetz thimble can serve as a perfect contour in the limit of $\hbar \rightarrow 0$. It is likely the case, but not to be proven here, that there exists a family of perfect contours $\gamma(\hbar)$ such that the thimble is obtained as the limit $\lim_{\hbar\rightarrow 0^+} \gamma(\hbar)$. On the other hand, due to the non-uniqueness of perfect contours (see Appendix~\ref{app:nonunique}), it is certainly not in general the case that every family of perfect contours has the Lefschetz thimble as its classical limit.

The situation is dramatically simplified in the case of one complex dimension. There the perfect contour is unique (at any given $\hbar$, provided it exists), and the Lefschetz thimbles are also uniquely identified (within the correct homology class) by obeying $\Im S = 0$. Therefore, in one complex dimension, the $\hbar \rightarrow 0$ limit of perfect contours is always a Lefschetz thimble, and every Lefschetz thimble may be so obtained.

\section{Two toy sign problems}\label{sec:toys}

\subsection{The cosine model}\label{sec:cosine}

We begin with a one-site model introduced in~\cite{Lawrence:2020irw}:
\begin{equation}\label{eq:cosine}
    \omega = (\cos \theta + \epsilon) d\theta
    \text.
\end{equation}
For real values $\epsilon \ge 1$, this model has no sign problem. In general we may take $\epsilon$ to be complex (as discussed in~\cite{Yamauchi:2021kpo}); however in this section we will restrict ourselves to the sign problem encountered when $\epsilon \in (-1,1)$.

\begin{figure}
    \centering
    \includegraphics[width=3in]{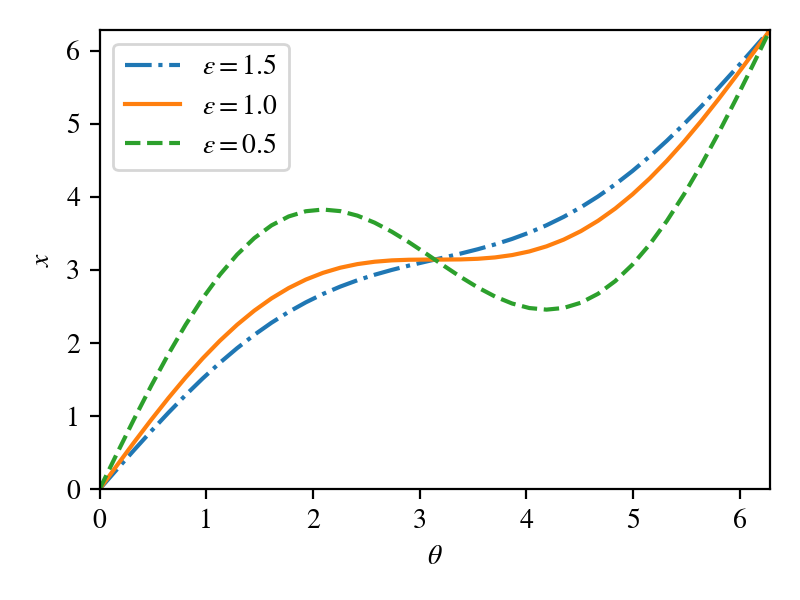}
    \caption{Behavior of the normalizing flow of the cosine model at $\epsilon=1.5, 1.0$ and $0.5$.\label{fig:cosine-nf}}
\end{figure}
A normalizing flow $\theta(x)$ for the cosine model is a 1-dimensional map from the distribution Eq.~(\ref{eq:cosine}) to the uniform distribution on the same domain. The inverse of the normalizing flow $x(\theta)$ can be written analytically by solving the following equation
\begin{equation}\label{eq:cosinenfdef}
    \frac{d\theta(x)}{dx}\;\frac{\cos(\theta(x))+\epsilon}{\epsilon} = 1
\end{equation}
and we find 
\begin{equation}
    x(\theta)=\left( \sin(\theta)+\epsilon\theta\right)/\epsilon \text.
\end{equation}
The normalizing flows for $\epsilon=1.5, 1.0, 0.5$ are shown in Figure~\ref{fig:cosine-nf}. For the cosine model with $\epsilon<1$, the distribution vanishes at two values of $\theta$, forcing $\frac{dx(\theta)}{d\theta}=0$ at those two $\theta$s for Eq.~(\ref{eq:cosinenfdef}) to hold. As a result, the normalizing flows are multi-valued.

Note that when $\epsilon$ is real, the integrand on the real line is also real, having no variations in phase apart from (when $\epsilon \in (-1,1)$) two isolated zeros. As a result, the argument in Section~\ref{sec:perfect} above indicates that the real line is at least a locally optimal contour, in the sense that no sufficiently small deviation from the real line can improve the sign problem.

In fact, for real $\epsilon \ge 0$, the real line is a globally optimal contour as well. This may be seen by examining the dependence of the form $\omega$ on the imaginary part of $\theta$. Writing $\theta = \phi + i \eta$ for $\phi,\eta \in \mathbb R$, the quenched Boltzmann factor is
\begin{equation}
|\omega| = \left(\cos\theta \cosh \eta + i \sin\theta \sinh \eta + \epsilon\right)|d\theta|
    \text.
\end{equation}
Focusing on the hyperbolic dependence on $\eta$, we see that any deformation from the real line will entail an increase in the quenched Boltzmann factor at every point, and therefore in the quenched partition function.

Following the strategy of Section~\ref{sec:bounds}, we may establish this fact algebraically. Define a new $1$-form
\begin{equation}
    \alpha = \frac 1 2 \left|\cos\Big(\frac{\theta + \bar\theta}{2}\Big) + \epsilon\right|(d\theta + d\bar\theta)
    \text.
\end{equation}
To see that $\alpha$ is closed, it is sufficient to note that the scalar piece depends only on $\Re \theta$, and the differential piece is $d \Re \theta$. Note that $\alpha$ is not exact: if it was, then it would integrate to $0$ and no non-trivial bound could be obtained.

The absolute value of this new form is a lower bound on that of the original form. This may be seen from the inequality $|d\theta + d\bar\theta| \le 2|d\theta|$ and the result of Appendix~\ref{app:cosine-bound} (for $\epsilon \le 1$):
\begin{equation}\label{eq:cosine-bound}
    \left|\cos\Big(\frac{\theta + \bar\theta}{2}\Big) + \epsilon\right| \le |\cos\theta + \epsilon|\text.
\end{equation}

As a result of $d\alpha = 0$, the integral is independent of the choice of contour $\gamma$ (as long as $\gamma$ is homologous to the real circle). We can therefore evaluate the integral along the real circle:
\begin{equation}
    Z' = \int_\gamma \alpha
    = \int_0^{2\pi} |\cos\theta + \epsilon| \,d \theta
    \text.
\end{equation}

Finally, we observe that $|Z'|$ provides a lower bound for the quenched partition function:
\begin{equation}
|Z'| = \left|\int_\gamma \alpha\right|
\le \int_\gamma |\alpha|
\le \int_\gamma |\omega| = Z_Q(\gamma)
\text.
\end{equation}
Since $Z'$ is exactly equal to the quenched partition function on the real circle, we have established that the real circle is globally optimal.

\subsection{The one-plaquette model}\label{sec:plaquette}
Our second toy model is the following differential form
\begin{equation}\label{eq:plaquette-form}
    \omega = e^{\beta \cos\theta} d\theta\;\text,
\end{equation}
which is a one-plaquette model of the lattice Abelian gauge theory. There is no sign problem when $\beta \in \mathbb R$; we will allow $\beta$ to be complex. The original field space $\theta$ is the real circle, and we parameterize its complexified space as $\theta=\phi+ i\eta$ with $\phi\in S^1$ and $\eta\in \mathbb{R}$. Following the procedure introduced in the previous sections, we establish a lower bound on the quenched partition of Eq.~(\ref{eq:plaquette-form}) by constructing a closed differential form $\alpha$ which satisfies $|\alpha| \le |\omega|$ and computing the absolute value of its partition function.  We start by introducing a new coordinate $\phi'(\eta, \phi)$ and writing the new form $\alpha$ as
\begin{equation}
    \alpha = c \min_{\eta}(|e^{\beta \cos(\phi(\phi',\eta)+i\eta)}|) \; d\phi'\;\text.
\end{equation}
Here, $\min_{\eta}$ means that the function is minimized by varying $\eta$ while $\phi'$ is held fixed. The scalar piece of $\alpha$ depends only on $\phi'$ and the differential piece is $d\phi'$, so $\alpha$ is a closed form. Since $\min_{\eta}(|e^{\beta \cos(\eta', \phi')}|) \le |e^{\beta\cos\theta}|$ for any fixed $\phi'$ by definition, the new differential form $\alpha$ satisfies $|\alpha| \le |\omega|$ when the constant $c$ is chosen such that $|c d\phi'| \le |d\theta|$. 
 
The choice of the coordinate $\phi'(\phi, \eta)$ determines the tightness of the lower bound we obtain. To find a useful $\phi'$, it is helpful to inspect the behavior of the absolute value of $\omega$ at $\eta\rightarrow \infty$ and $-\infty$:
\begin{equation}
    \frac{|\omega|}{|d\theta|} = \exp\left(\Re\beta\sin\phi\cos\eta + \Im\beta\sin\phi\sinh\eta   \right)\;\text.
\end{equation}
For example, in cases $\Re\beta>0, \Im\beta > 0$, the behavior of the form at $\eta\rightarrow\infty$ is 
\begin{align}
    \frac{|\omega|}{|d\theta|} \rightarrow 
    \begin{cases}
    \infty & \tan^{-1}\left(-\gamma\right)< \phi < \tan^{-1}\left(-\gamma\right)+\pi\\
    -\infty & \mathrm{else}
    \end{cases}
\end{align}
with $\gamma=\Re\beta/\Im\beta$. The asymptotic behavior at $\eta\rightarrow-\infty$ is
\begin{align}
    \frac{|\omega|}{|d\theta|} \rightarrow 
    \begin{cases}
    \infty & \tan^{-1}\left(\gamma\right) -\pi < \phi < \tan^{-1}\left(\gamma\right)\\
    -\infty & \mathrm{else}
    \end{cases}\;\text.
\end{align}
 The inverse tangent function is defined to take a value in the range $\left[-\frac{\pi}{2}, \frac{\pi}{2}\right]$. This asymptotic behavior makes it clear that defining $\phi' = \phi$ will not give a non-trivial bound ---  for example, $\alpha=0$ for all $\phi'$ when $\Re\beta=0$. A useful change of coordinate is such that the line of fixed $\phi'$ connects the diverging ($|\omega|\rightarrow\infty$) region at $\eta=\pm\infty$ and likewise for the vanishing ($|\omega|\rightarrow0$) region. On such a line, the new differential form $\alpha$ takes a non-zero value when its asymptotic behavior is divergent, therefore the bound on $Z_Q$ will be non-zero. Noting that the range of $\phi$ where $|\omega|$ diverges at $\eta\rightarrow\infty$ is shifted by
\begin{equation}
 s = \tan^{-1}\left(-\frac{\Re\beta}{\Im\beta}\right)+\pi - \tan^{-1}\left(\frac{\Re\beta}{\Im\beta}\right)
\end{equation}
in the positive $\phi$ direction with respect to such a range at $\eta\rightarrow-\infty$, one such change of coordinates is 
\begin{equation} \label{eq:plaq_phip}
    \phi = \phi' + \frac{s}{\pi}\;\tanh^{-1}(\eta)\;\text.
\end{equation}
% such that 
% \begin{equation}
%     \theta = \phi' + \frac{s}{\pi}\;\tanh^{-1}(\eta) + i\eta
% \end{equation}
Note that the scalar part of $\alpha$ is periodic with $\phi'$. 

Following the choice of $\phi'$ in Eq.~(\ref{eq:plaq_phip}), the constant $c$ must be chosen to guarantee $|c d\phi'| \le |d\theta|$. Parameterizing $d\theta = d\phi+i d\eta =\cos\xi+i\sin\xi$ with $-\pi \le \xi < \pi$ such that $|d\theta|=1$, $\phi'$ gives
\begin{equation}
    |d\phi'| = \left|\cos\xi - \frac{s}{\pi(1+\eta^2)} \sin \xi\right|
    \le \sqrt{1+\left( \frac{s}{\pi} \right)^2}\;\text.
\end{equation}
Therefore 
\begin{equation}
    c = \frac{1}{\sqrt{1+\left( \frac{s}{\pi} \right)^2}} 
\end{equation}
satisfies $|d\phi'| < |d\theta|$ for any $\xi$. Putting all elements together, the differential form is
\begin{equation}
    \alpha = \frac{\min_{\eta}(|e^{\beta \cos(\phi'+\frac{s}{\pi}\tanh^{-1}\left(\eta\right) + i\eta)}|)}{\sqrt{1+\left( \frac{s}{\pi} \right)^2}} \; d\phi'\text.
\end{equation}
For $\beta$ with $\Re\beta<0, \Im\beta>0$, the differential form $\alpha$ can be constructed in the same way. Starting by the shifting constant
\begin{equation}
 s_- = \tan^{-1}\left(\frac{\Re\beta}{\Im\beta}\right)+\pi - \tan^{-1}\left(-\frac{\Re\beta}{\Im\beta}\right)
\end{equation}
and introducing the new coordinate 
\begin{equation}
    \phi = \phi' - \frac{s_-}{\pi}\;\tanh^{-1}(\eta)\;\text,
\end{equation}
the differential form is
\begin{equation}
    \alpha = \frac{\min_{\eta}(|e^{\beta \cos(\phi'-\frac{s_-}{\pi}\tanh^{-1}\left(\eta\right) + i\eta)}|)}{\sqrt{1+\left( \frac{s_-}{\pi} \right)^2}} \; d\phi'\text.
\end{equation}
We do not address the region of $\beta$ with $\Re\beta < 0$ because of the symmetry of the model, that is $\omega(\beta, \theta) = \omega(-\beta,\theta+\pi)$.

\begin{figure}
    \centering
    \includegraphics[width=0.45\textwidth]{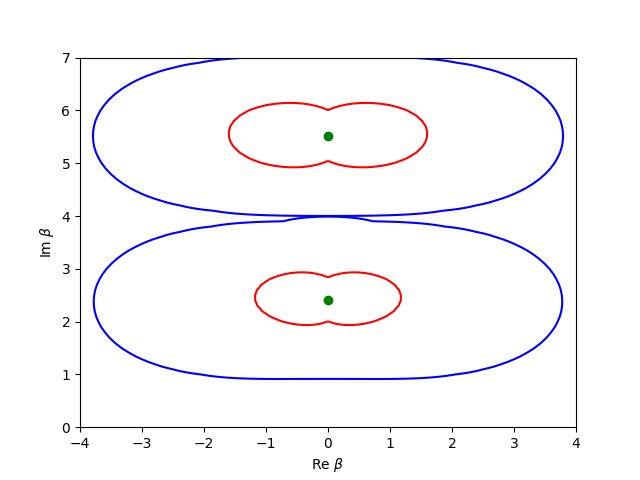}
    \caption{The availability and non-availability of perfect contours for the one-plaquette model. Perfect contours provably do not exist for the regions of $\beta$ closed by the red lines. For the ranges of $\beta$ outside the blue lines, prefect contours with the average phase $\ge 0.999$ were constructed. Zeros of the partition function are marked in green. Note that because the form optimization is done over a restricted set of differential forms, the red bounds are not tight.\label{fig:plaquette}}
\end{figure}

The upper bound on the average phase of the one plaquette model is derived by taking the ratio of the partition function $Z$ and the integral of $\alpha$. Using the fact that $\phi'=\phi$ and $d\phi'=d\phi$ along the real circle, the integral can be computed as
\begin{equation}
    Z' = \int_{-\pi}^{\pi} \frac{\min_{\eta}(|e^{\beta \cos(\phi'+\frac{s}{\pi}\tanh^{-1}\left(\eta\right) + i\eta)}|)}{\sqrt{1+\left( \frac{s}{\pi} \right)^2}} \; d\phi'
\end{equation}
for $\Re\beta,\Im\beta>0$ and likewise for the cases $\Re\beta<0, \Im\beta>0$. In FIG.~\ref{fig:plaquette}, we show in red line the values of $\beta$ where the bound on the average phase $|Z/Z'|$ is 1. The partition function vanishes at $\beta=2.404, 5.520$, and the bounds prove that perfect contours do not exist for the region closed by the red lines. The figure also shows the regions where perfect contours exist. We constructed the perfect contours by optimizing the Fourier series
\begin{equation}
    \theta = \phi + i\left( a_0 + \sum_{j=0}^{n} c_j \cos(j\phi) + s_j\sin(j\phi) \right)
\end{equation}
with $n=10$. For the region outside the blue lines, we could find contours of integration with the average phase $\ge 0.999$ via the Fourier series above. As a consequence of the strong duality between our bounds and the contour optimization problem, we know that the gap between the blue and red lines is due to some combination of the looseness of the bounds or the restrictive search of perfect contours via the Fourier series. In a larger-scale search for contour deformations, this gap may be taken as a sign that a more intensive search may be profitable.

\section{Abelian Yang-Mills}\label{sec:yang-mills}

We will now use the results of the previous section to establish exponential best-case scaling for the average sign in lattice Abelian Yang-Mills, across a wide range of complex couplings. The lattice model being considered has action
\begin{equation}
S = \beta \sum_r \left[1 - \cos(\theta_{r,1} + \theta_{r+\hat x,2} - \theta_{r+\hat y,1} - \theta_{r,2})\right]
\text,
\end{equation}
where $\theta_{r,i}$ denotes the value of $\mathfrak{u}(1)$ associated to the gauge link from site $r$ to site $r + \hat i$, and $\beta$ is the complex coupling.

We will restrict ourselves to two-dimensional lattices with open boundary conditions (or at least two-dimensional lattices with trivial top homology group). Such systems may be exactly solved by performing a gauge transformation that decouples gauge degrees of freedom. The partition function can be written (up to irrelevant overall normalization) as a product over the $N$ original plaquettes~\cite{Migdal:1975zg,Gross:1980he}:
\begin{equation}
    Z = \left[\int_0^{2\pi} d\theta e^{-\beta \cos \theta}\right]^N
    \text.
\end{equation}
As a result, the properties of two-dimensional Abelian Yang-Mills are determined by the toy integral of Section~\ref{sec:plaquette}. As we will see, this extends to contour deformations.

First, let $\gamma_1$ be a contour deformation for the single-plaquette theory. This may be extended naturally to a contour $\gamma = (\gamma_1)^{\times N}$ for use in the $N$-plaquette theory. The quenched partition function of this contour is $Z_Q(\gamma) = Z_Q^{(1)}(\gamma_1)^N$; as a result the average sign obtained with this family of contours scales exactly exponentially with $N$.

Now let $\gamma_1^*$ be the optimal contour for the one-plaquette theory. It is not \emph{a priori} obvious that following the construction above to obtain $\gamma^* = (\gamma^*_1)^{\times N}$ yields an optimal contour for the theory with $N$ plaquettes. It is conceivable, for example, that although the theory itself is local (in the sense that correlations decay at large distances), the optimal contour is not. In fact, we can use the results of the previous sections to show that this is not the case.

Let $\alpha_1$ be a closed differential form dominated by $\omega_1 = e^{\beta \cos \theta} d \theta$; in other words, a differential form that yields a lower bound on the quenched partition function of the one-plaquette model. Just as a contour for the $N$-plaquette model can be constructed as $N$ copies of a single-plaquette contour, we can construct a differential form $\alpha_N$ by wedge product:
\begin{equation}
    \alpha_N(\theta_1,\ldots,\theta_N) = \bigwedge_n \alpha_1(\theta_n)
    \text.
\end{equation}
A wedge product of closed differential forms $\alpha_1$ is also closed, and it is similarly easy to verify that it is dominated by the integrand of the $N$-plaquette model. The lower bound on the quenched partition function obtained in this fashion again scales exactly exponentially with volume:
\begin{equation}
    \left|\int \alpha_N\right| = \left|\int \alpha_1\right|^N\text.
\end{equation}

Analysis of the one-plaquette model therefore yields both lower and upper bounds on the optimal contour of the $N$-plaquette model. Moreover, by strong duality, the two bounds coincide. Therefore we see that the factorized contour constructed as above is in fact optimal (although not necessarily unique). This result is general enough to hold for any theory which exactly factorizes.

We conclude by noting, as mentioned in Section~\ref{sec:nf}, perfect contours can be obtained from the analytic continuation of normalizing flows. In the case of Yang-Mills in two dimensions, normalizing flows may be easily constructed~\cite{Kanwar:2021tkd}, although it appears that not much is known about their analytic continuation. For the sake of brevity we will not discuss this connection further.

\section{Further discussion}\label{sec:discussion}
The central acheivement of this work is the construction in Section~\ref{sec:bounds} of a general method for proving lower bounds on the quenched partition function, independent of what integration contour is used. Sufficiently strong bounds of this form would constitute a no-go theorem for the use of contour deformations alone to cure a sign problem. Additionally, as seen in Section~\ref{sec:yang-mills}, bounds of this form are naturally exponential in spacetime volume. There is no apparent obstacle to applying this method to non-trivial lattice field theories beyond the solvable models considered in this paper.

In Section~\ref{sec:convex-optimization} we established that the optimization is such bounds is strongly dual to the optimization of (generalized) contours. That is, we showed that if the best possible bound that can be obtained by the methods of Section~\ref{sec:bounds} is that $Z_Q(\gamma) \ge Z^*_Q$, then there is in fact some contour deformation $\gamma$ that obtains a quenched partition function of $Z^*_Q$. This result has two immediate consequences. First, as a philosophical matter, a failure of the contour deformation method for a particular theory can always be explained using closed-form-induced bounds. Second, going forward, searches for good contours of integration can in principal be augmented by simultaneous searches for closed differential forms. The searches can be terminated when the contour optimization approaches the bound optimization; conversely if a search is unable to bring the optimum and the bound close to each other, that is an indication that a larger ansatz may be required. This provides a ``stopping criterion'' in the training of a contour which has so far been unavailable.

Note that although there are two convex optimization problems discussed---that of optimizing the bounds and that of optimizing generalized contours---the task of optimizing contour deformations themselves is not convex, or at least has not been shown to be.

We discussed briefly in Section~\ref{sec:thimbles} the relation between locally perfect contours and the Lefschetz thimbles. The fact that the thimbles form the $\hbar \rightarrow 0$ limit of perfect contours is suggestive: it indicates that locally perfect contours may share other properties in common with thimbles even away from the classical limit. For example: the thimbles form a linearly independent basis for homology. We have established that locally perfect contours form a basis for homology, but establishing that this basis is linearly independent (or finding a counterexample) remains an open problem. (Some recent related discussion, regarding quantum corrections to Lefschetz thimbles, may be found in~\cite{Gantgen:2023byf}.)

Finally, the behavior of perfect contours in the limit of large $\hbar$ has not yet been determined, beyond the observation of Section~\ref{sec:thimbles} that the imaginary part of the Jacobian must be exactly constant.

\begin{acknowledgements}
We are indebted to Frederic Koehler for discussions on the conditions for strong duality and a pointer to Sion's minimax theorem. We are furthermore grateful to Gurtej Kanwar for comments on an earlier draft of this manuscript. S.L.~was supported at the beginning of this work by the U.S.~Department of Energy under Contract No.~DE-SC0017905, and subsequently by a Richard P.~Feynman fellowship from the LANL LDRD program.  Y.Y.~is supported by the U.S. Department of Energy grant No.~DE-FG02-00ER41132. This work was additionally supported by the U.S. Department of Energy through the Los Alamos National Laboratory. Los Alamos National Laboratory is operated by Triad National Security, LLC, for the National Nuclear Security Administration of U.S. Department of Energy (Contract No. 89233218CNA000001).
\end{acknowledgements}

\appendix

\section{Perfect contours are not unique}\label{app:nonunique}
In this appendix we show, via a counterexample in the Gaussian case, that perfect contours (satisfying $Z_Q = |Z|$) are in general not unique. Note that this property is special to multi-dimensional integrals.

Consider a lattice action with two degrees of freedom:
\begin{equation}
    S = \frac 1 2 \left(z_1^2 + z_2^2\right)\text.
\end{equation}
This is a solvable theory, equivalent (up to a change of variables) to free scalar field theory on a lattice with two sites. The real plane is a perfect contour, but it is not unique. Let $x \in \mathbb R^2$, and parameterize a contour by $z = M x$ for a $2\times 2$ complex matrix $M$ of the form
\begin{equation}
    M = \left(\begin{matrix}
        \sqrt{1 - \mu^2} & i \mu\\
        -i \mu & \sqrt{1 - \mu^2}
    \end{matrix}\right)\text,
\end{equation}
where $\mu$ lies in the interval $(-1,1)$. As $\det M = 1$, the effective action in terms of the parameterizing variable $x$ is $S_{\mathrm{eff}}(x) = \frac 1 2 x^T M^T M x$. The mass matrix evaluates to the identity $M^T M = I$,
and so the effective action is real for all $x$. Note that the action obeys $S \rightarrow +\infty$ as $x \rightarrow \infty$ as long as $|\mu| < 1$.

Even this one-parameter family of perfect contours is not exhaustive. For example, it is easy to find perfect contours in this case that are not flat; that is, for which the Jacobian is not constant (although its determinant is). We accomplish this by allowing the rotation parameter $\mu$ to vary as a function of $|x|^2$. Define a contour by
\begin{equation}\label{eq:big-perfect-family}
\left(
\begin{matrix}
    z_1\\
    z_2
\end{matrix}
\right)
     = \left(\begin{matrix}
            1 & i \mu(|x|^2)\\
        -i \mu(|x|^2) & 1
        \end{matrix}
    \right) 
\left(
\begin{matrix}
    x_1\\
    x_2
\end{matrix}
\right)
    \text.
\end{equation}
The Jacobian of the parameterization is
\begin{equation}
    J =
    \left(
    \begin{matrix}
        1 & i \mu + 2 i \mu'x_2\\
        i \mu + 2 i \mu' x_1 & 1
    \end{matrix}
    \right)
\end{equation}
where both $\mu$ and its derivative $\mu'$ are implicitly functions of $|x|^2$. It is apparent that for any function $\mu(|x|^2)$ obeying
\begin{equation}
(\mu + 2 \mu' |x|)^2 < 1
\text,
\end{equation}
we have a perfect contour given by Eq.~(\ref{eq:big-perfect-family}).

\section{Derivation of Eq.~(\ref{eq:cosine-bound})}\label{app:cosine-bound}
Writing $\theta = \phi + i \eta$ for real $\phi,\eta$, and assuming $\epsilon \in [0,1]$, we will show that
\begin{equation}
    0 \le |\cos\phi\cosh\eta - i \sin\phi\sinh\eta + \epsilon|^2 - |\cos\phi + \epsilon|^2
\end{equation}
which is equivalent to Eq.~(\ref{eq:cosine-bound}). Expanding all terms yields
\begin{align}
    0 \le &\cos^2\phi (\cosh^2\eta - 1) + \sin^2\phi\sinh^2 \nonumber\\
    &+ 2 \epsilon \cos\phi(\cosh\eta -1) \eta\text.
\end{align}
The first and second terms are combined by noting that $1 = \cosh^2\eta - \sinh^2\eta$. Additionally using the identity $\sinh^2 \eta = \frac 1 2 (\cosh 2\eta - 1)$ we find that the above inequality is equivalent to
\begin{equation}
    0 \le \frac 1 2 (\cosh 2 \eta - 1) + 2 \epsilon \cos\phi(\cosh \eta -1)\text.
\end{equation}
As $|\epsilon \cos\phi| \le 1$, the above is implied by
\begin{equation}
    \frac{\cosh 2 \eta - 1}{\cosh \eta - 1} \ge 4\text,
\end{equation}
which can be readily checked.

\bibliography{refs}

\end{document}